\documentclass[conference]{IEEEtran}
\usepackage{amssymb,amsmath,epsfig,latexsym,graphicx,bm,nccmath}
\usepackage{mathtools}

\usepackage{epstopdf}
\usepackage{cite}
\usepackage{amsfonts}
\usepackage[ruled, lined, linesnumbered, commentsnumbered, longend]{algorithm2e}
\usepackage{textcomp}
\usepackage{xcolor}
\def\BibTeX{{\rm B\kern-.05em{\sc i\kern-.025em b}\kern-.08em
		T\kern-.1667em\lower.7ex\hbox{E}\kern-.125emX}}
\usepackage{caption}	
\usepackage{subcaption}
\usepackage[utf8x]{inputenc}
\usepackage{etoolbox}

\usepackage[skip=0pt]{caption}
\usepackage[font=footnotesize,skip=0pt]{subcaption}
\usepackage{easyReview} 

\title{Joint User Association and Bandwidth Assignment for Digital Twin-Assisted Multi-RAT Networks}

\author{
	\IEEEauthorblockN{Manobendu Sarker\IEEEauthorrefmark{1}, Md. Zoheb Hassan\IEEEauthorrefmark{2}, and Georges Kaddoum\IEEEauthorrefmark{1}}
	\IEEEauthorblockA{\IEEEauthorrefmark{1}{Department of Electrical Engineering}, \'{E}cole de Technologie Sup\'{e}rieure, Universit\'{e} du Qu\'{e}bec, Montr\'{e}al, QC,  Canada \\
		Emails: \{manobendu.sarker, georges.kaddoum\}@etsmtl.ca}
	\IEEEauthorblockA{\IEEEauthorrefmark{2}{Department of Electrical and Computer Engineering}, Universit\'{e} Laval, Qu\'{e}bec City, QC, Canada \\
		Email: md-zoheb.hassan@gel.ulaval.ca}
}

\makeatletter
\patchcmd{\@maketitle}
{\addvspace{0.5\baselineskip}\egroup}
{\addvspace{-1\baselineskip}\egroup}
{}
{}
\makeatother

\begin{document}
	
	\maketitle
	
	\begin{abstract}
		In this paper, we investigate user equipment (UE)-radio access technology (RAT) association and bandwidth assignment to maximize sum-rates in a multi-RAT network. To this end, we formulate an optimization problem that jointly addresses UE association and bandwidth allocation, adhering to practical constraints. Because of the NP-hard nature of this problem, finding a globally optimal solution is computationally infeasible. To address this challenge, we propose a centralized and computationally efficient heuristic algorithm that aims to maximize sum-rates while enhancing quality of service (QoS). Yet, the proposed approach requires global channel state information (CSI) for near-optimal performance, which incurs substantial overhead and data collection costs in large-scale multi-RAT networks. To alleviate this burden, we use a digital twin (DT) of the multi-RAT network, leveraging its context-awareness to acquire global CSI with reduced overhead. Our numerical results reveal that our approach improves sum-rates by up to 43\% over baseline method, with less than a 5\% deviation from the theoretical optimal solution, while achieving up to a 43\% improvement in QoS. Further analysis reveals that our method not only surpasses the optimal solution in terms of QoS enhancement, but also ensures significant computational efficiency.
	\end{abstract}
	
	\vspace{-2mm}
	\section{Introduction}
	The growing demand from data-intensive applications such as augmented reality (AR), virtual reality (VR), and the Internet of Things (IoT) underscores the limitations of single RAT networks in meeting the diverse needs of modern user equipment (UE). In this context, a multi-RAT system, which integrates technologies like 5G New Radio (NR) and WiFi 6, offers a promising solution, as it enables seamless connectivity across densely populated and high-mobility environments \cite{10054381}. As each RAT has unique capabilities in coverage, capacity, and latency, effective resource management (e.g., UE-RAT association and bandwidth assignment) is crucial for balancing load, reducing interference, and enhancing throughput \cite{andrews2014will}. Specifically, an optimal assignment of UEs to RATs based on factors like signal quality, load, and UE requirements is important to prevent congestion \cite{attiah2020survey}. In addition, bandwidth allocation further ensures that UEs achieve required data rates, particularly in the constrained settings of dense networks \cite{galanopoulos2020multi}. Therefore, efficient resource management can significantly improve the quality of service (QoS) in a multi-RAT system, making it highly adaptable to diverse UE demands and traffic patterns.
	
	To date, several previous studies investigated UE-RAT association and bandwidth assignment in multi-RAT networks. For instance, Attiah et al. \cite{attiah2020survey} surveyed UE association and spectrum sharing for millimeter wave systems, proposing solutions for seamless connectivity. Similarly, in \cite{galanopoulos2020multi}, the authors examined spectrum reallocation for higher data rates and improved quality of experience (QoE). While these studies considered single-RAT association per UE, another pertinent investigation \cite{shi2017dual} focused on multiple associations to enhance throughput and offloading performance.
	
	However, despite recent advances, available multi-RAT resource management approaches frequently lack scalability and computational efficiency, as well as rely on precise system dynamics and global channel state information (CSI). Gathering global CSI incur high overhead in large and heterogeneous networks with fast-changing channels. Another limitation of past works is that most solutions overlook the capacity limits of RAT-supported base stations (BSs)/access points (APs)—an assumption that, if adequately addressed, could improve signal detection and reduce inter-UE interference. Furthermore, prior studies also largely overlook the balance between maximizing sum-rates and ensuring QoS, i.e., the probability of UEs meeting minimum rate requirements. Seeking to address these limitations, in this study, we propose efficient radio resource management strategies for multi-RAT networks.
	
	In recent years, achieving real-time situational awareness in dynamic wireless networks has garnered significant research attention, particularly through digital twin (DT) technology. DT replicates physical wireless networks in a virtual space using realistic models and data from the physical environment \cite{DT_1, DT_3, DT_4}. DT offers continuous environmental sensing, high-fidelity 3D ray tracing, and robust data processing capability for dynamic wireless networks \cite{DT_2}. Such attributes can enable multi-RAT networks to acquire real-time information on network CSI, available bandwidth, and UE QoS requirements  with a significantly reduced direct interaction with the physical network. Accordingly, in this study, we explore a DT-assisted multi-RAT system combining 5G NR and WiFi 6 technologies and introduces a novel joint UE-RAT association and bandwidth allocation scheme. The contributions of this paper can be summarized as follows.
	
	\smallskip \noindent $\bullet$ We present a DT model that enables an efficient collection of global CSI for a multi-RAT network. To the best of our knowledge, this is the first application of DT in a multi-RAT network that provides a viable approach for global CSI collection while minimizing information acquisition costs.
	
	\smallskip \noindent $\bullet$ We formulate an NP-hard problem focused on maximizing network sum-rates by optimizing UE-RAT association and bandwidth assignment, subject to constraints on bandwidth, minimum rate, multi-connectivity, and UE association limits for each RAT-supported BS/AP.
	
	\smallskip \noindent $\bullet$ To address this computationally intensive problem, we propose an efficient UE-RAT association and bandwidth assignment scheme that achieves near-optimal sum-rate performance, enhances QoS, and maintains computational efficiency.
	
	\smallskip \noindent $\bullet$ The results of numerical evaluations demonstrate that the proposed scheme achieves near-optimal solution for maximizing sum-rates and improved QoS with significantly lower computational costs as compared to the theoretical optimal solution.	
	\begin{figure}[tb]	
		\centering
		\includegraphics[width=0.9\linewidth, draft=false]{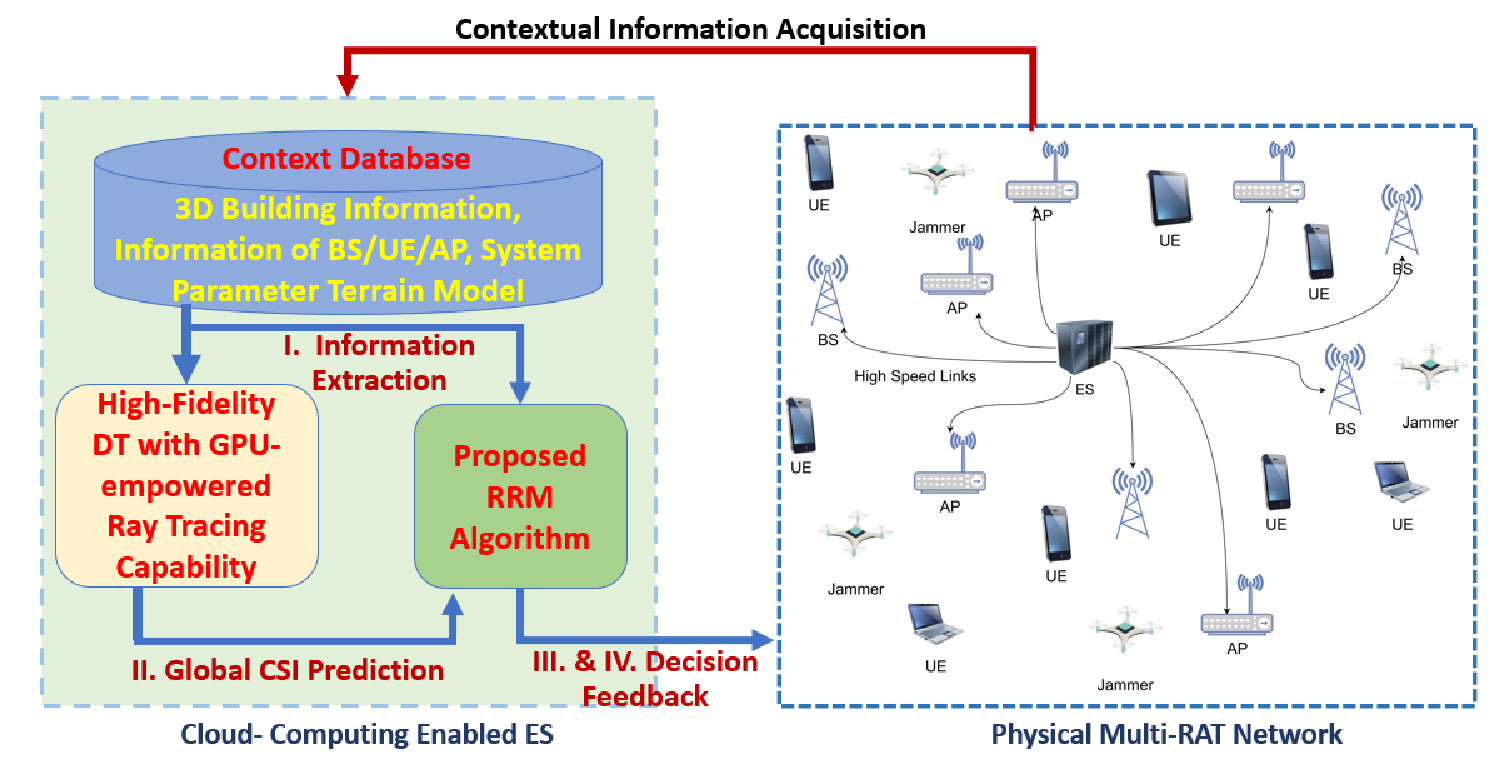}
		\caption{Overall architecture of the proposed multi-RAT system.}
		\label{architecture}
	\end{figure}
	 \vspace{-2mm}
	\section{System Overview}
	We consider a downlink multi-RAT network system with 5G NR and WiFi 6 technologies,  where $B$ BSs using 5G NR and $A$ APs using WiFi 6 serve multiple single-antenna UEs (see in Fig. \ref{architecture}). Each BS/AP is equipped with $M \geq 1$ antennas, which may vary between RATs. All BSs/APs and $U$ UEs are randomly distributed across the service area, and each UE supports multi-mode, multi-band operations, thus enabling seamless mobility between RATs. The overall system is centrally managed by a cloud-computing (CC) empowered edge-server (ES), which supports fast and reliable control information exchanges with  all BSs/APs via wired feedback links. Using a DT and resource control unit, the ES  dynamically updates UE-RAT associations and bandwidth assignments based on the context feedback from the network. In addition, $L$ UAV-mounted jammers, each with $N$ antennas, are randomly placed within the service area. Jamming source locations and interference levels are assumed to be known to the ES, facilitated by its jamming detection algorithms. For the sake of simplicity, we treat jammer signals as a form of external interference, leaving the impact and mitigation of various types of jamming attacks for future work. Let $\mathcal{B}, \mathcal{A}, \mathcal{U}$, and $\mathcal{L}$ denote the sets of BSs, APs, UEs, and jammers, respectively. {Throughout the paper, we refer to the collection of all BSs and APs as service points (SPs), denoted by $\mathcal{J} = \mathcal{B} \cup \mathcal{A}$.}
	
	The envisioned system has various forms of interference, including cross-RAT, intra-RAT inter-SP, and intra-SP multi-UE interference. Since managing such vast ranges of interference is beyond the scope of this work, we consider that the two RATs operate on different frequency bands to eliminate the cross-RAT interference which aligns with typical deployments. In addition, we adopt the orthogonal multiple access approach used in the existing standards, where each UE associated with a given SP is allocated unique sub-channels. This allocation eliminates multi-UE interference within an SP. However, since UEs support multi-mode and multi-band operations, they can simultaneously associate with multiple SPs over different frequency channels. Hence, on each frequency channel, the UE can receive co-channel interference from neighboring SPs who also transmit to their associated UEs on the same frequency channel. In order to mitigate such intra-RAT inter-SP interference and to maximize system throughput, the resource control unit of ES periodically optimizes UE-RAT associations and bandwidth assignments.
	
	 \vspace{-2mm}
	\section{Description of the Digital Twin Network}
	Optimal UE-RAT association and bandwidth assignment require extensive global CSI acquisition across the multi-RAT network, specifically necessitating $M \times B \times A$ CSI estimations per UE. In large-scale networks, collecting such extensive CSI with limited pilot resources is challenging and incurs high overhead. In this context, due to its rich multi-modal sensing and environmental awareness, DT can accurately track UEs virtually and estimate both large-scale and small-scale fading coefficients \cite{Channel_Twin}. For instance, a high-fidelity DT can use a 3D environmental map and the electromagnetic (EM) properties of objects to estimate CSI \cite{Ahmed_1, Ahmed_2}. A proof-of-concept demonstration of DT-enabled CSI acquisition was previously reported in \cite[Fig. 3]{Colosseum}. Thus, DT offers a feasible technology to acquire the global CSI  of multi-RAT networks while minimizing the signaling overhead and cost of information collection from the physical network.
	
	In this study, we assume a CC-enabled ES hosts a high-fidelity DT of the multi-RAT network, equipped with real-time 3D ray-tracing and synchronized with the physical network through appropriate interfaces. The DT maintains a virtual network topology, providing thus a digital abstraction of nodes such as BSs, APs, UEs, and jammers, including key attributes such as locations, physical dimensions, and radio characteristics (e.g., antenna models, bandwidth, transmit power, etc.). It also integrates a high-resolution 3D environmental map, as shown in \cite{Ahmed_1}, to position network nodes. In addition, the ES hosts a context database with information on factors such as weather, building geolocations and EM properties, dynamic UE positions, and other parameters, all of which are regularly updated from the physical network.
	
	\textbf{A Walk-through:} As shown in Fig. \ref{architecture}, the proposed framework consists of the following four key steps.\\ 
	\noindent $\bullet$ \textbf{Step 1:} The ES retrieves necessary information for CSI estimation from the context database. \\
	\noindent $\bullet$ \textbf{Step 2:} The ES updates the DT with the latest contextual information and calculates channel coefficients using 3D ray-tracing.\\
	\noindent $\bullet$ \textbf{Step 3:} Using the updated channel coefficients and other parameters to the resource control unit, the ES determines UE-RAT associations and allocates bandwidth. \\
	\noindent $\bullet$ \textbf{Step 4:} The ES communicates the updated resource allocation decisions to the physical multi-RAT networks (i.e., SPs), thus enabling data transmission based on these optimized allocations. 
	
	In what follows, we describe the DT modeling parameters.  The proposed DT modeling parameters are modularized, allowing them to be adapted to various environments.
	\vspace{-3mm}
	\subsection{DT Modeling Parameters}
	\subsubsection{SP-UE Channel Model}
	We consider that the UEs experience both line-of-sight (LOS) and non-line-of-sight (NLOS) reflections when they receive signals from SPs. Accordingly, the channel vector between UE $i$ and SP $j$, $\textbf {h}_{ij} \in \mathcal{C}^{M \times 1}$ can be modeled as shown in Eq. (\ref{eq.1}) \cite{10149096}.
	\begin{equation}
				\footnotesize
		\textbf {h}_{ij} = g_{0} \textbf{b} \left ({{{\alpha _{0}},{\beta _{0}}} }\right) + \sqrt {\frac {1}{Q}} \sum_{q = 1}^{Q} g_q \textbf{b} \left ( \alpha_q,\beta_q \right), \label{eq.1}
	\end{equation}
	where $Q$ is the number of NLOS paths, $g_0$ and $g_q$ denote the large-scale fading coefficients of the LOS path and the $d$-th NLOS path, respectively, and $g_q \sim \mathcal{CN}(0, 10^{-\text{PL}_{ij}/10})$ for $q = 0,1,...,Q$. $\text{PL}_{ij}$ is the path loss in dB, defined as shown in Eq. (\ref{eq.2}) \cite{9473488}. 
	\begin{equation}
		\text{PL}_{ij} = \text{PL}_{FS}(d_0) + 10n \ \text{log}_{10}(d_{ij}/d_0) + E, \label{eq.2}
	\end{equation}
	where $d_{ij}$ is the distance between UE $i$ and SP $j$, $\text{PL}_{FS}(d_0)$ is the free space path loss at reference distance $d_0 = 1m$, $n$ represents the path loss exponent, and $E$ denotes the log-normal shadowing with 0 mean and $\sigma^2$ standard deviation. Next, $\alpha_q$ and $\beta_q$ for $q = 0,1,...,Q$ are the the vertical and horizontal angles of departure (AoD), respectively, of the LoS/NLoS path. Finally, $\textbf{b} \left ( \alpha_q,\beta_q \right)$ denotes the steering vector, which is given by $\textbf{b} \left ( \alpha_q,\beta_q \right) = [1,e^{j(2\pi\delta/\lambda) \ \text{sin} \ \alpha \ \text{cos} \ \beta},...,e^{j(2\pi\delta(M-1)/\lambda) \ \text{sin} \ \alpha \ \text{cos} \ \beta}]^T$, where $\delta$ is the antenna spacing and $\lambda$ is the signal wavelength. 
	\subsubsection{Jammer-UE Channel Model}
	In our system, UAV-mounted jammers disrupt SP-UE transmissions by emitting jamming signals, with UAVs in motion.  The signal received by a UE from a jammer comprises LOS, strong NLOS reflections, and multiple scattered components, all of which generate multi-path fading \cite{feng2006path}. The likelihood of receiving LOS and strong NLOS components is considerably higher than that of fading effects. Accordingly, small-scale fading effects are disregarded. The channel between UE $i$ and jammer $l$, $\textbf{h}_{il} \in \mathcal{C}^{N \times 1}$, is given by Eq. (\ref{eq.3}) \cite{10149096}. 
	\begin{equation} 
		\footnotesize
		{{\textbf {h}}_{il}} = {g_{0}}{\textbf {c}}\left ({{{\theta _{0}},{\varphi _{0}}} }\right) + \sqrt {\frac {1}{Z}} \sum \limits _{z = 1}^{Z} {{g_{z}}{\textbf {c}}\left ({{{\theta _{z}},{\varphi _{z}}} }\right)}, \label{eq.3}
	\end{equation}
	where $Z$ is the number of NLOS paths; $\theta_z$ and $\varphi_z$ denote the horizontal and vertical AoDs of each path, respectively; $\textbf{c}(\theta_z, \varphi_z)$ is the steering vector; and  $g_z \sim \mathcal{CN}(0, 10^{\text{PL}_{il}/10})$, $\forall z$, with $\text{PL}_{il}$ as the jammer-UE path loss. Given UAV's mobility, 
	the modified path loss between UE $i$ and jammer $l$ can be expressed as shown in Eq. (\ref{eq.4}) \cite{7842372}.
	\begin{align}
			\footnotesize
		\text{PL}_{il} &= \text{PL}_{FS}(d_0) + 10n \ \text{log}_{10}(d_{ij}/d_0) - 10 \text{log}_{10}(\Delta h/h_{\text{opt}}) \nonumber\\ 
		&+ 10 \text{log}_{10}c_{\mathrm{p}} + 10e \text{log}_{10}((f_{\mathrm{e}}+\Delta f)/f_{\mathrm{e}}) + E, \label{eq.4}
	\end{align}
	where $\Delta h = |h_{\text{gnd}} - h_{\text{opt}}|$; $h_{\text{gnd}}$ is the receiver’s height, and $h_{\text{opt}}$ is the UE height that minimizes path loss. Constant loss factor $C_{\text{p}} = 10 \ \text{log}_{10} c_{\text{p}} \geq 0$ dB accounts for foliage and antenna orientation losses on the UAV. The Doppler shift $\Delta f = (\Delta v/c)f_{\text{e}}$, with $\Delta v$ as UAV-UE relative speed and $f_{\text{e}}$ the emitted frequency, modifies the observed frequency $f = f_{\text{e}} + \Delta f$. For small velocities, the term $10 e \ \text{log}_{10}((f_{\mathrm{e}}+\Delta f)/f_{\mathrm{e}})$ is negligible.  
	\subsubsection{Model for Signal Transmission and Data Rate}
	In the proposed system, each UE can simultaneously receive downlink data from multiple SPs over different frequency channels. Let  $\mathcal{V}_i$ be the subset of SPs serving the $i$-th  UE, $s_{ij}$ be the transmitted signal the $j$-th SP to the $i$-th UE with $\mathrm{E}{|s_{ij}|^2} = 1$, and  $s_{il}$ be signal transmitted from the $l$-th jammer to the $i$-th UE with unit average power. The received signal at the $i$-th UE can be expressed as shown in Eq. (\ref{received_signal}).
		\begin{equation}
			\footnotesize
			\begin{aligned}
				\mathbf{y}_i =  \sum_{j \in \mathcal{V}_i} \Bigg( \sqrt{\eta_{ij}} \mathbf{h}_{ij} s_{ij}  
				+ \sum_{p \in \mathcal{J} \setminus j} \sqrt{ \eta_{ip}} \mathbf{h}_{ip} s_{ip} 
				+ \sum_{l \in \mathcal{L}} \sqrt{ \eta_{il}} \mathbf{h}_{il} s_{il} + w_i \Bigg),
			\end{aligned}
			\label{received_signal}
		\end{equation}
	where  $\eta_{ij}$ is the transmit power of the data signal $s_{ij}$ with $0 < \sum_{i \in J}\eta_{ij} \leq P_j$; $\eta_{il}$ is the transmit power of the jamming signal $s_{il}$ with $0 < \sum_{i \in J}\eta_{il} \leq P_l$; and $w_i = \mathcal{CN}(0, \sigma_i^2)$ represents additive white Gaussian noise (AWGN) with a mean of zero and variance $\sigma_i^2$. Here,  $P_j$ and $P_l$ are the maximum transmit power of SP $j$ and jammer $l$, respectively. In Eq.  \eqref{received_signal}, the first, second, and third terms represent the expected data signal of UE $i$ from the associated SP, received interference from neighbor SP of the associated SP, and jamming interference, respectively.
	
	The downlink data rate achieved at the $i$-th UE is given by 
	\begin{equation}
		R_{i} = \sum_{j \in \mathcal{V}_i} \ x_{ij} y_{ij} \text{log}_2 \left(1+ \Gamma_{ij} \right), \label{rate}
	\end{equation}
	where $x_{ij} \in \{0,1\}$ indicates the binary UE association variable. Here, $x_{ij} = 1$ signifies that the $j$-th UE is assigned to $j$-th BS(AP) of 5G-NR(WiFi) RAT, and $x_{ij} = 0$  otherwise. In addition, $y_{ij}$ indicates the bandwidth assignment to UE $i$ at the $j$-th associated SP. Assuming that the $i$-th UE is associated with the $j$-th SP, its received signal-to-interference-plus-noise ratio (SINR) is expressed  based on Eq. (\ref{received_signal}) as shown below.
	\begin{equation}
		\footnotesize
		\Gamma_{ij} = \frac{ \eta_{ij} \|\mathbf{h}_{ij}\|^2 }{\sum_{p \in (\mathcal{B}/\mathcal{A}) \setminus j}   \eta_{ip} \|\mathbf{h}_{ip}\|^2+ \sum_{l \in \mathcal{L}}  \eta_{il} \|\mathbf{h}_{il}\|^2+\sigma_i^2}. \label{SINR}
	\end{equation}
	
	\section{Problem Formulation}
	Our objective is to maximize the overall system rate during the downlink phase through UE-RAT association  and bandwidth assignments under different constraints. The problem can be formulated for a given power allocation as follows:
	
		{\begin{subequations}
			\footnotesize
			\begin{align}
				\textbf{P0}: {\underset{x_{ij} \in \{0,1\}, y_{ij} > 0} {\operatorname{max}}}  & \ \ \sum_{i \in \mathcal{U}} \ R_i,  \label{obj}   \\
				\textrm{s.t.} \quad &  \ \sum_{j \in \mathcal{V}_i}  x_{ij} y_{ij} \leq W_j, \ \forall i, \label{bandwidth}\\
				& \ \sum_{i \in \mathcal{U}}  x_{ij}  \leq \overline{J}_j, \ \forall j, \label{limit_AP}\\
				& \ \sum_{j \in \mathcal{J}}  x_{ij}  \leq \underline{J}, \ \forall i, \label{limit_UE}\\
				&  \sum_{j \in \mathcal{V}_i} x_{ij} y_{ij}  \text{log}_2 \left(1+\Gamma_{ij}\right) \geq \overline{R}_i, \ \forall i, \label{QoS}
			\end{align}
		\end{subequations}		}
	where the SINR $\Gamma_{ij}$ is defined in Eq. (\ref{SINR}).
	
	The constraint defined in Eq. (\ref{bandwidth}) limits the total bandwidth resources allocated at SP $j$ to its serving UEs to $W_j$. Furthermore, constraint (\ref{limit_AP}) ensures that the total number of UEs associated with SP $j$ does not exceed $\overline{J}_j$. Note that the values of $W_j$ and $\overline{J}_j, \ \forall j$, can vary for different RATs. We also account for UE multi-connectivity by allowing a UE to associate with multiple SPs ($\underline{J} \geq 1$) through constraint (\ref{limit_UE}), thus providing SPs with greater flexibility in data transmission to meet UE rate requirements. Subsequently, the minimum rate requirement for a UE $i$ from its associated SPs and corresponding bandwidth assignments, specified to be greater than $\overline{R}_i$, which pertains to QoS, is enforced through constraint (\ref{QoS}). Of note, the value of $\overline{R}_i, \ \forall i$, may vary across UEs, depending on different requested services. {Note that the optimization variables, $x_{ij}$ and $y_{ij}$, in problem $\textbf{P0}$ are binary and positive continuous, respectively.
	}
	
	In problem \textbf{P0}, the product of $x_{ij}$ and $y_{ij}$ in the objective function (\ref{obj}) and constraint \eqref{bandwidth}, non-linear SINR functions in the constraint (\ref{QoS}), and combinatorial constraints introduce non-convexity. Particularly, the well-known multi-dimensional knapsack problem, also NP-hard, can be reduced to problem \textbf{P0}, making it a computationally intractable NP-hard problem. The proof is omitted here due to space constraints.

	\vspace{-1mm}
	\section{Proposed UE-RAT Association and Bandwidth Assignment Scheme}
	\subsection{Proposed Solution}
	To address the high computational cost of solving problem \textbf{P0} in dynamic wireless networks, we propose a two-phase heuristic algorithm for UE-RAT association  and bandwidth assignment while striking a suitable balance between optimality and computational complexity. In the initial phase of the proposed scheme, initial associations are first established between UEs and SPs, forming the basis for bandwidth assignment, while the second phase adjusts these allocations to maximize the sum-rate. In what follows, we detail each phase, along with their respective computational complexities.
	
	\subsubsection{Initial Assignment Phase}
	\label{Initial Assignment Phase}
	The first phase aims to (i) assign each UE to appropriate SPs under constraints (\ref{limit_AP}) and (\ref{limit_UE}) and (ii) allocate initial bandwidth based on these associations, respecting constraint (\ref{bandwidth}). This phase is based on SINR values and has the following two stages:
	
	\smallskip \noindent $\bullet$	 \textbf{Stage I. UE-RAT Association  Assignment}: For each UE $i$, all available SPs are sorted by SINR in the descending order. The UE is assigned to SP(s) iteratively, starting from the highest SINR SP from the sorted list, while ensuring compliance with constraint (\ref{limit_AP}). If the SP cannot support additional UE due to its existing UE associations, the next SP in the sorted list is considered. This process continues until each UE has $\underline{J}$ SP associations, meeting thus constraint (\ref{limit_UE}), and the UE association  matrix $\mathbf{x}$ is updated accordingly.
	
	\smallskip \noindent $\bullet$	 \textbf{Stage II. Bandwidth Assignment}: At this stage, our scheme assigns bandwidth to each UE based on UE association  results from the previous stage. Specifically, each SP allocates bandwidth proportionally to each UE’s SINR relative to the total SINR of its associated UEs, thereby ensuring compliance with constraint (\ref{bandwidth}), and consequently updates $\mathbf{y}$ accordingly. For a UE $i$ under SP $j$ (where $x_{ij} = 1$), bandwidth $y_{ij}$ is given by Eq. \eqref{weigth_SINR}.
	\vspace{-2mm}
	\begin{equation} 
				\footnotesize
		y_{ij} = W_j \times \frac{\Gamma_{ij}}{\sum_{k \in \mathcal{M}_{j}} \Gamma_{ik}}, \label{weigth_SINR} 
	\end{equation} 
	where $W_j$ is the SP $j$'s maximum bandwidth, and $\mathcal{M}_{j}$ is the set of UEs associated with SP $j$. 
	
	\subsubsection{Refined Assignment Phase}  
	\label{Refined Assignment Phase}
	In the second phase, the scheme iteratively refines initial UE associations and bandwidth assignments so as to maximize sum-rates, with options to satisfy each UE's minimum rate requirement (i.e., QoS) as needed. During each iteration, for a given SP $j$, the scheme identifies its associated UEs. For each UE $i$, it then ranks all other SPs (except for $j$ and any SPs currently associated with UE $i$) based on SINR in the descending order. From this list, the scheme selects the SP that maximizes the sum-rate, with an option to meet UE $i$'s minimum rate requirement, $\overline{R}_i$, as applicable. To limit the search space, only the top $l_m \times J$ SPs from the list are considered, where $0 < l_m < 1$. 
	
	Algorithm \ref{alg:selection} details the SP selection process for UE $i$. For each SP in the sorted list, lines 2-12 compute the sum-rate difference, $\Delta S$, and the status $F \in \{0,1\}$, indicating whether UE $i$’s current rate meets or exceeds $\overline{R}_i$ due to the temporary association of SP $j$ to UE $i$. Furthermore, line 13 identifies the SP $j$ that maximizes the combined value of $\Delta S$ and $F$. Of note, Algorithm \ref{alg:selection} is sufficiently flexible to improve sum-rates while optionally ensuring that UE $i$'s rate, $R_i$, meets or exceeds $\overline{R}_i$, based on the value of $\zeta$. To focus solely on improving sum-rates, $\zeta$ is set to 0; conversely, setting $\zeta = 1$ aims to enhance sum-rates while ensuring $R_i \geq \overline{R}_i$. Therefore, based on the value of $\zeta$, the algorithm makes the final selection of SP $j^*$ for UE $i$ in line 13, and both $\mathbf{x}$ and $\mathbf{y}$ are updated accordingly in line 14. This adjustment process (Algorithm \ref{alg:selection}) repeats until the change in sum-rates between consecutive iterations falls below a small threshold $\epsilon$.
	\begin{algorithm}[htbp]
				\footnotesize
		\caption{SP Selection Procedure for UE $i$}
		\label{alg:selection}
		\SetAlgoNlRelativeSize{-1}
		\KwIn{Sorted list of SPs, $\{\overline{R}_i\}$, $\{{W}_j\}$, $\{{\Gamma_{ij}}\}$, $\zeta$}
		\KwOut{Updated $ \mathbf{x} $ and $ \mathbf{y} $}
		\textbf{Initialize} $\Delta S \gets 0_{1 \times J}$, $ F \gets 0_{1 \times J}$;\\
		\For{$j \in $ the sorted list}{
			\If{SP $j$ can associate with UE $i$ according to constraint (\ref{limit_AP})}{
				Allocate bandwidth according to Eq. (\ref{weigth_SINR})\;
				$\Delta S_j \ \gets$ Calculate the difference in sum-rates\;
				\uIf{$R_i \geq \overline{R}_i$}{
					$F_j \ \gets$ 1\;
				}
				\Else{
					$F_j \ \gets$ 0\;
				}
			}
		}
		$j^* \gets {\underset{}{\operatorname{argmax}}} \ (\Delta S + F \times \zeta )$;\\
		Assign SP $j^*$ for UE $i$ and update $ \mathbf{y} $ accordingly;
	\end{algorithm}	
	\vspace{-2mm}
	\subsection{Computational Complexity of the Proposed Scheme}
	The first stage’s computational complexity, due to searching across SPs for all UEs, is $\mathcal{O}(JU)$, which is similar for the \textbf{Stage II}. Hence, the total complexity of the initial assignment phase is $\mathcal{O}(2JU)$. Next, the computational complexity of selecting a suitable SP for a UE is $\mathcal{O}(l_m J)$. Accordingly, the complexity of processing all UEs associated with a SP in one iteration is $\mathcal{O}(|\mathcal{M}_j| l_m J^2)$, where $|\mathcal{M}_j| \ll U$. With $\log(1/\epsilon)$ iterations for convergence, the overall complexity of the refined assignment phase is $\mathcal{O}(|\mathcal{M}_j| l_m J^2 \log(1/\epsilon))$. Therefore, the total computational complexity of the proposed UE-RAT association and bandwidth assignment scheme is $\mathcal{O}(2JU + |\mathcal{M}_j| l_m J^2 \log(1/\epsilon))$.
	
	\vspace{-2mm}
	\section{Numerical Analysis}
	In this section, we numerically evaluate the performance of proposed scheme in a multi-RAT network with the presence of aerial jammers. The network operates over a 1000 m$^2$ area with $J = 40$ SPs: 20 5G NR BSs and 20 WiFi 6 APs serving $U$ single-antenna UEs. Each BS has $M_{\text{5G NR}} = 32$ antennas with a transmit power limit of $P_{\text{5G NR}} = 100$ W, while each AP is equipped with $M_{\text{WiFi 6}} = 8$ antennas and a power cap of $P_{\text{WiFi 6}} = 40$ W. The BSs and APs can serve up to 20 and 8 UEs, respectively. Bandwidth constraints are set at $W_{\text{5G NR}} = 100$ MHz for each BS and $W_{\text{WiFi 6}} = 80$ MHz for each AP, with carrier frequencies of $f_{\text{5G NR}} = 25$ GHz and $f_{\text{WiFi 6}} = 6$ GHz, respectively \cite{9473488} . The system also includes $L = 10$ UAV-mounted jammers, each with $N = 4$ antennas and a transmit power limit of $P_{\text{Jammer}} = 20$ W. The noise power is set to $10^{-9}$ W. The SPs, UEs, and jammers are randomly placed within the area, maintaining the minimum 100 m distance between same-type SPs. Channel parameters follow those previously outlined in \cite{10149096}. Uniform power allocation is applied with each UE subject to a minimum rate requirement of $\overline{R} = 0.5$ Mbps. Finally, to ensure convergence, the threshold for the proposed scheme is set to $\epsilon = 0.001$.
	
	\subsubsection{Comparison of Average Sum-rate Performance}
	\label{Comparison of Average Sum-rate Performance}
\begin{figure}[tb]	
		\centering
		\includegraphics[scale=0.3]{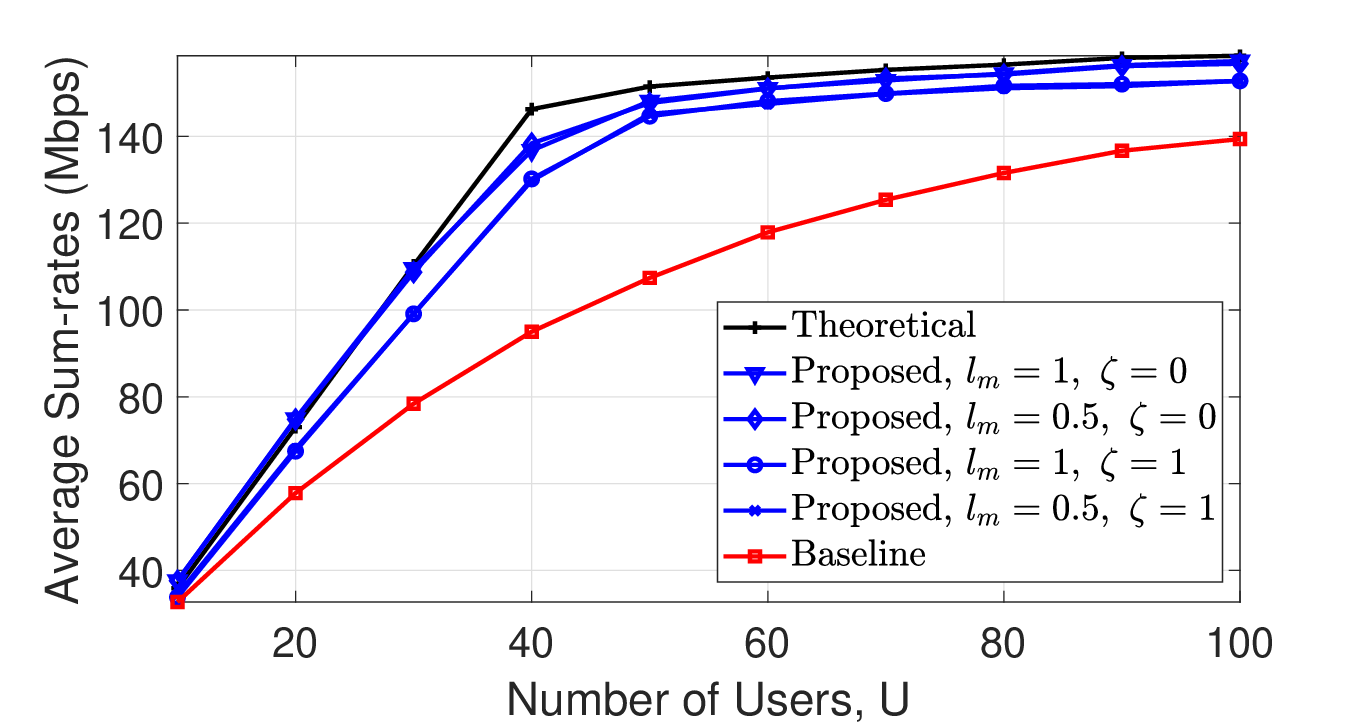}
		\caption{Average sum-rate performance of UE association and bandwidth schemes for varying $U$ with $\overline{J} = 1$.}
		\label{sumrate_association_bandwidth}
	\end{figure}

	Fig. \ref{sumrate_association_bandwidth} shows the average sum-rate performance of our proposed scheme along with the `Theoretical' solution from the Gurobi solver \cite{gurobi_mip_complexity} and a `Baseline' scheme that combines the initial UE association described in Section \ref{Initial Assignment Phase} with a Round-Robin algorithm-based bandwidth assignment. The results reveal that theoretical solution consistently achieves the highest sum-rate, with the proposed scheme performing within $5\%$ of this optimal level when $\zeta = 0$,  demonstrating thus near-optimal performance due to its adaptive UE association and bandwidth adjustments aimed solely at maximizing sum-rate. However, setting $\zeta = 1$ increases the performance gap to up to $10\%$, which still remains near-optimal. By contrast, the baseline scheme significantly underperforms, showing a decrease of up to $43\%$ and $37\%$ for $\zeta = 0$ and $\zeta = 1$, respectively, due to its limited bandwidth flexibility. Importantly, performance under $l_m = 1$ and $l_m = 0.5$ in the proposed scheme is similar, indicating that reducing the search space does not impact sum-rate performance (see in Section \ref{Comparison of Computational Complexity Performance} for further discussion). At higher UE densities, sum-rate gains diminish due to RAT bandwidth constraints.
	
	\subsubsection{Comparison of Average Success Probability Performance}
	\label{Comparison of Average Success Probability Performance}
\begin{figure}[tb]	
		\centering
		\includegraphics[scale=0.3]{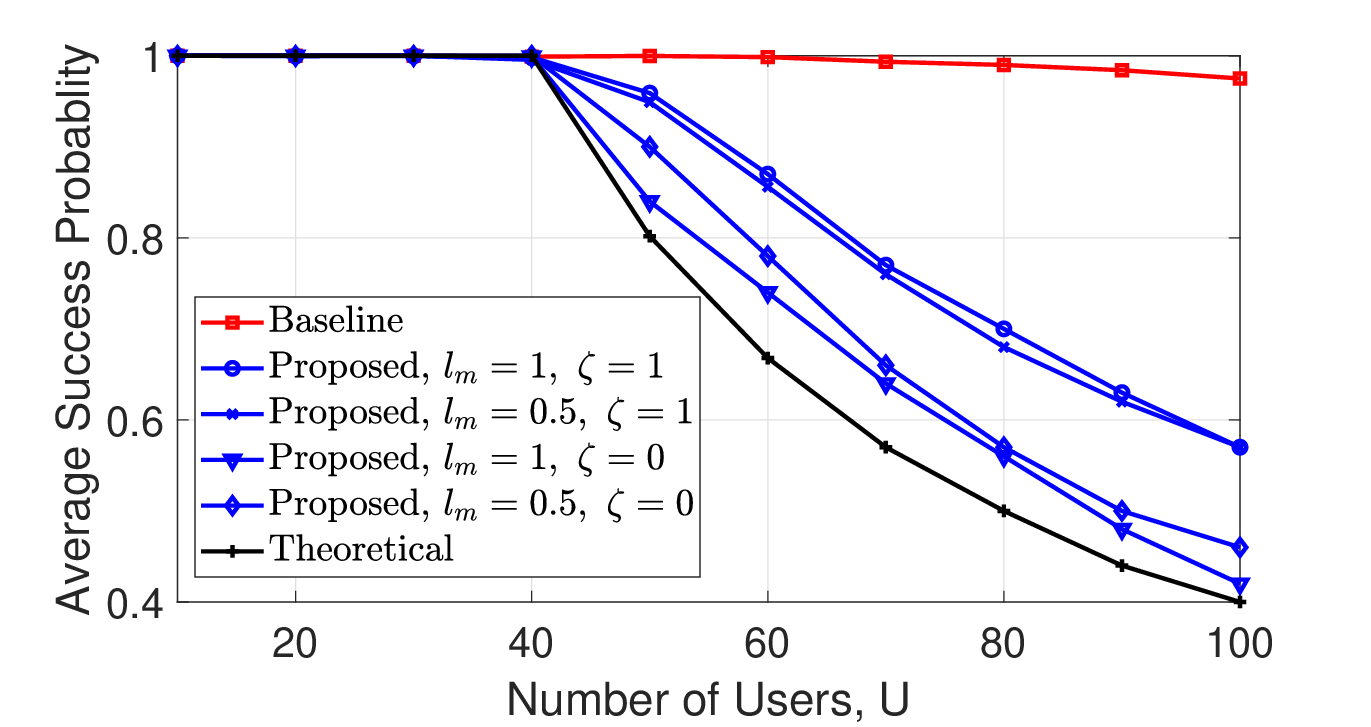}
		\caption{Average success probability of different joint UE association and bandwidth schemes for varying $U$ with $\overline{J} = 1$.}
		\label{success_association_bandwidth}
	\end{figure}

	Fig. \ref{success_association_bandwidth} shows the average success probability across the considered schemes, defined as the percentage of UEs meeting their minimum rate requirement, i.e., ensuring QoS. The results show that the baseline scheme achieves the highest success probability, while the theoretical solution has the lowest due to its primary focus on maximizing sum-rate. The proposed scheme provides up to a $43\%$ increase in success probability over the theoretical solution for $\zeta = 1$, improving thus the sum-rate while ensuring QoS during SP selection and bandwidth allocation (see in Section \ref{Refined Assignment Phase}). However, the success probability decreases to $17\%$ when $\zeta = 0$, since Algorithm \ref{alg:selection} then prioritizes sum-rate maximization exclusively. Both $l_m = 1$ and $l_m = 0.5$ configurations perform similarly, with $l_m = 0.5$ offering a more computationally efficient solution.
	
	\subsubsection{Comparison of Computational Complexity Performance}
	\label{Comparison of Computational Complexity Performance}
\begin{figure}[tb]	
		\centering
		\includegraphics[scale=0.3]{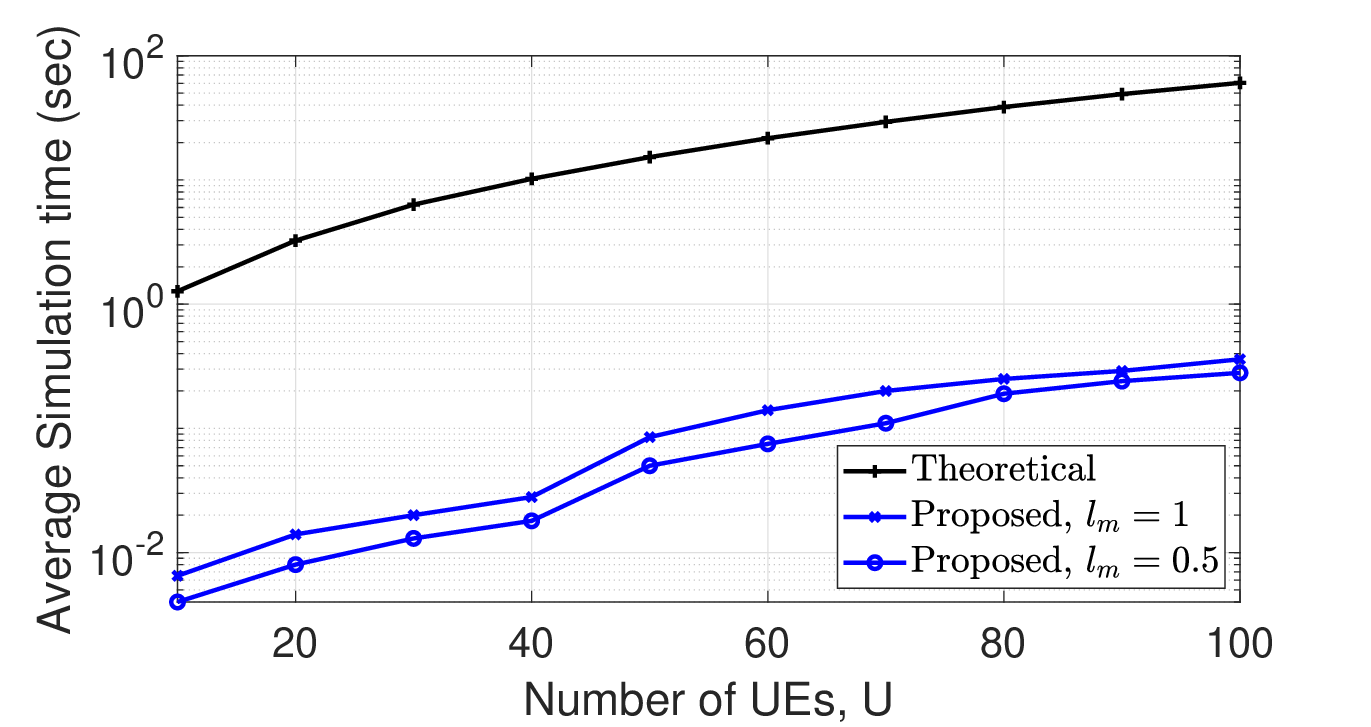}
		\caption{Average simulation time of theoretical and heuristic schemes for varying $U$ with $\overline{J} = 1$.}
		\label{time_association_bandwidth}
	\end{figure}

	The Gurobi solver’s computational complexity for solving problem \textbf{P0} is $O\left(2^{JU} \left((3J + 2U)^3 + (3J + 2U) \times (2JU)^2\right)\right)$ \cite{gurobi_mip_complexity}, \cite{bertsimas1997introduction}. To compare, the proposed scheme’s computational complexity (Section \ref{Refined Assignment Phase}) is significantly lower by up to two and three orders in $J$ and $U$, respectively. Fig. \ref{time_association_bandwidth} confirms this conclusion, showing up to $566\%$ and $363\%$ reductions in runtime for $l_m = 0.5$ and $l_m = 1$, respectively, as compared to the theoretical solution. Therefore, the $l_m = 0.5$ option offers the shortest execution time with minimal sum-rate impact, making it a favorable choice. 
	
	\subsubsection{Effect of Multi-Connectivity}
\begin{figure}[tb]	
		\centering
		\includegraphics[scale=0.3]{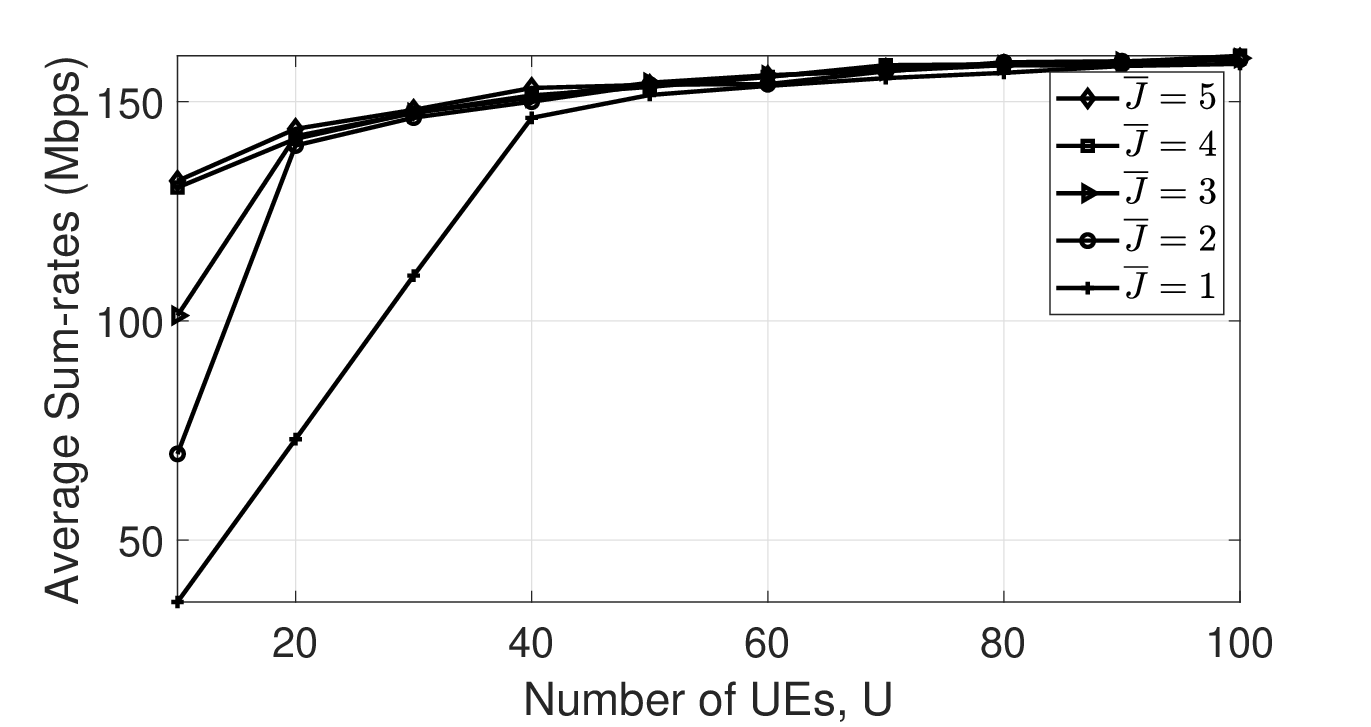}
		\caption{Average sum-rate performance with varying $\overline{J}$ and $U$.}
		\label{association_effect}
	\end{figure}

\begin{figure}[t]	
		\centering
		\includegraphics[scale=0.3]{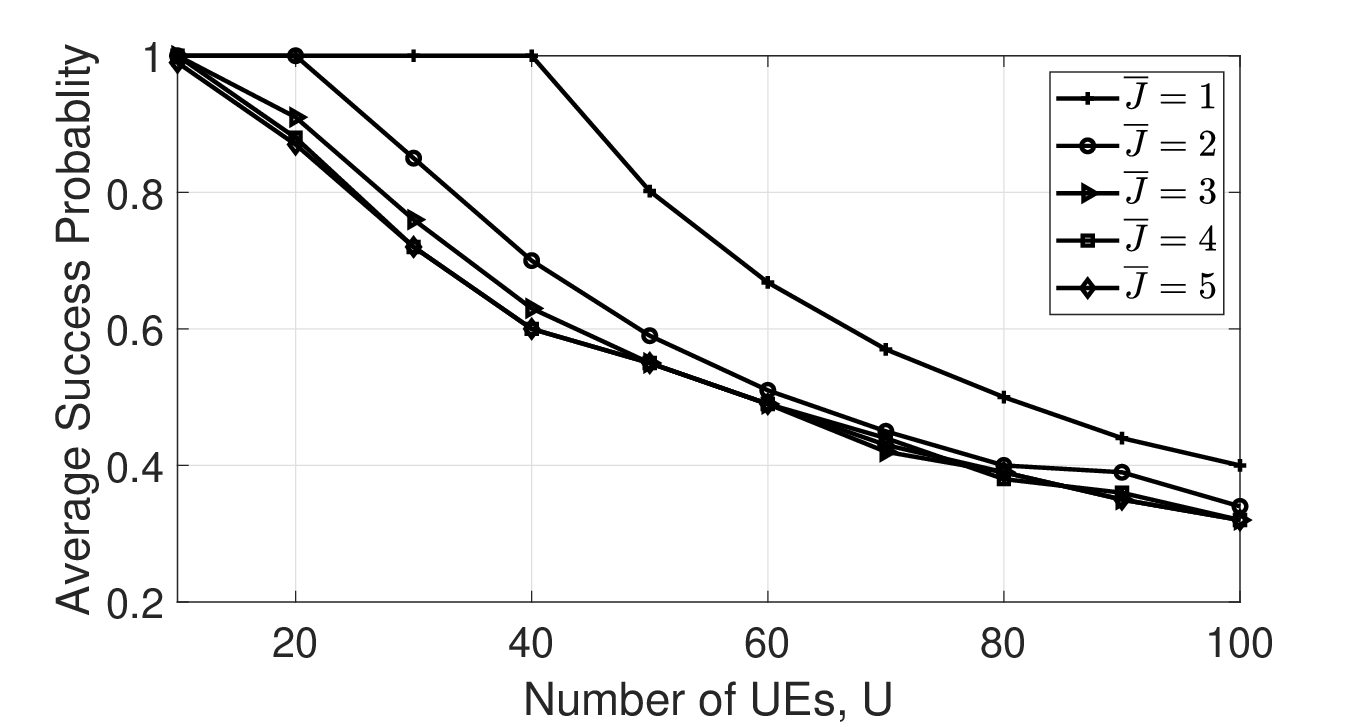}
		\caption{Average success probability performance with varying $\overline{J}$ and $U$.}
		\label{association_effect_success}
	\end{figure}

	Fig. \ref{association_effect} shows the impact of multi-connectivity (parameter $\overline{J}$) on sum-rate. The theoretical solution serves as a benchmark to evaluate the effects of multi-connectivity. As can be seen in Fig. \ref{association_effect}, increasing $\overline{J}$ boosts sum-rate, especially at lower UE densities; however, performance saturates at higher densities, as, with increased $\overline{J}$, the bandwidth allocated per UE is reduced. Moreover, as shown in Fig. \ref{association_effect_success}, lower $\overline{J}$ values yield higher minimum success probabilities due to the problem \textbf{P0} objective. Conversely,  increasing $\overline{J}$ offers minimal success probability gains for high UE densities. Overall, lower UE densities benefit from higher $\overline{J}$ for sum-rate, while lower $\overline{J}$ balances sum-rate and success probability at high densities. Said differently,  the benefit of multi-RAT connectivity is diminished at high UE density due to bandwidth constraint per SP. Consequently, expanding RAT bandwidth is essential to improve sum-rate at high UE density.
	
	\section{Conclusion}
	In this paper, for maximizing sum-rates while improving the QoS rate for UE, we presented a computationally efficient centralized heuristic scheme for the joint UE-RAT association and bandwidth assignment problem. To facilitate centralized operation without incurring high overhead and information gathering costs, we proposed a DT model as a viable option to obtain global CSI. The results of our numerical analyses revealed that the proposed scheme achieves a favorable balance between computational efficiency, sum-rate, and QoS performance for multi-RAT networks, which makes it as a robust alternative to both the optimal solution and baseline approach. Furthermore, our simulation results indicated that expanding RAT bandwidth is critical for enhancing sum-rates, especially at higher UE densities. In future research, it would be necessary to explore the ways to improve the capability of the current multi-RAT optimization framework by integrating power allocation and further situational-awareness in DT. 
	\bibliography{reference}
	\bibliographystyle{ieeetr}
	
\end{document}